\documentclass[aps,prb,reprint,groupedaddress]{revtex4-1}
\usepackage{graphicx}  

\begin{document}


\title{
Particle-hole Asymmetry of Fractional Quantum Hall States in the Second Landau Level of a Two-dimensional Hole System}

\author{A. Kumar}
\author{N. Samkharadze}
\author{G.A. Cs\'{a}thy}
\email[]{gcsathy@purdue.edu} 
\affiliation{Department of Physics, Purdue University, West Lafayette, Indiana 47907, USA } 

\author{M.J. Manfra}
\affiliation{Department of Physics, Birck Nanotechnology Center, School of Materials Engineering, 
and School of Electrical and Computer Engineering,
Purdue University, West Lafayette, Indiana 47907, USA}

\author{L.N. Pfeiffer }
\author{K.W. West}
\affiliation{Department of Electrical Engineering, Princeton University, Princeton, New Jersey 08544, USA} 


\date{\today}

\begin{abstract}
We report the first unambiguous observation of a fractional quantum Hall state in the Landau level of
a two-dimensional hole sample at the filling factor $\nu=8/3$. We identified this state by a quantized Hall resistance and an activated temperature dependence of the longitudinal resistance and found an energy gap of 40~mK. To our surprize the
particle-hole conjugate state at filling factor $\nu=7/3$ in our sample does not develop down to 6.9~mK. 
This observation is contrary to that in electron samples in which the 7/3 state is typically more stable than the 8/3 state.
We present evidence that the asymmetry between the 7/3 and 8/3 states in our hole sample is due to Landau level mixing.
\end{abstract}

\maketitle

In a two-dimensional electron system (2DES) subjected to a perpendicular magnetic field $B$ the Coulomb interaction between the charge carriers leads to the emergence of prototype many body ground states unknown in any other condensed matter system. Well known examples are the series of fractional quantum Hall states (FQHS) of the lowest Landau level (LL) \cite{Tsui} developing at Landau level filling factors $\nu$ of the form
$m/(2m \pm 1)$, where $m$ is an integer. Extensive experimental and theoretical investigations 
\cite{Tsui99} have established that the parent FQHS are described by Laughlin's wavefunction \cite{Laughlin}
while the series of FQHS of the lowest LL can be described in the framework 
of Jain's weakly interacting composite fermion model \cite{JainCF}.

FQHS also form in the second LL (i.e. $2<\nu<4$) but, 
in contrast to their lowest LL counterparts, the nature of these states is
not well understood. Of these the $\nu=5/2$ even denominator FQHS has attracted a lot of attention 
\cite{willett87,pan99,eisen02,miller07,dean08,pan08,nuebler10,zhang10,csa10,xia04,g05,dean08-2,dolev08,choi08,willett09,xia10,dolev11} 
as it is believed to arise from a $p$-wave pairing of composite fermions described by the 
Moore-Read Pfaffian wavefunction \cite{moore91}. 
With increasing sample quality an increasing number of odd denominator FQHS  
have been observed in the second LL \cite{willett87,pan99,eisen02,miller07,dean08,pan08,nuebler10,zhang10,csa10,xia04,g05,dean08-2,dolev08,choi08,willett09,xia10,dolev11}. 
For the $\nu=7/3$ FQHS, the most prominent of these states, 
recent numerical work finds evidence of Laughlin correlations 
\cite{toke05,peters08-2,papic09}. Other authors find, however, the $\nu=7/3$ FQHS
to be either exotic, with a wavefunction that is {\it distinct} from Laughlin's wavefunction
\cite{read99,bonderson08,simion08}, or on the borderline between the Laughlin and exotic non-Abelian states \cite{wojs09}.
Results of recent energy gap measurements \cite{csa10} and of experiments probing the back-propagating neutral modes \cite{dolev11}
for the $\nu=7/3$ and its particle-hole conjugate $8/3$ FQHS in high density 2DES are
consistent with these states being of the Laughlin type.
Experiments on lower density 2DES in tilted magnetic fields, however, yielded surprising and yet
unexplained dependence of the energy gap at $\nu=7/3$ on the in plane magnetic field \cite{dean08-2,xia10}.
The nature of the odd denominator FQHS in the second LL remains yet to be ellucidated.

FQHS can be probed by varying the Landau level mixing (LLM) \cite{yoshi86}.
Since at large magnetic fields the cyclotron energy greatly exceeds the Coulomb energy,
the excited Landau levels can be neglected and the energy gap of the FQHS is therefore
solely determined by the Coulomb energy.
At low magnetic fields at which the Coulomb energy exceeds the cyclotron energy,
the gap is in infuenced by the higher Landau levels and therefore
mixing of the Landau levels due to the Coulomb energy has to be considered \cite{yoshi86}.
In the lowest LL, LLM is known to reduce the energy gaps of the FQHS \cite{yoshi86,melik97,murthy02}. 
In contrast, in the second LL LLM is not yet fully understood but it
is expected to have a more profound effect. 
Theoretical work on the $\nu=5/2$ Pfaffian found that LLM can lift the degeneracy of the Pfaffian and 
its non-equivalent particle-hole conjugate anti-Pfaffian 
\cite{levin07,lee07,wan08,wang09,peterson08,wojs10,bishara09,rezayi11},
it may induce a transition from the Pfaffian to the
anti-Pfaffian state \cite{levin07,lee07,wan08,wang09},
or it may enhance the $\nu=5/2$ Pfaffian \cite{wojs10}. Alternative
possibilities are a linear superposition of the Pfaffian and anti-Pfaffian and
spacially randomized domains of Pfaffian and anti-Pfaffian controlled by the disorder \cite{peterson08}.
Similar ideas should also apply for exotic odd denominator states in the second LL
which are degenerate at vanishing LLM \cite{read99,bonderson08,bernevig08,levin09}.

We have studied the FQHS of the second LL at extremely large LLM which is realized in a 
two-dimensional hole sample (2DHS). Indeed, due
to the larger effective mass of holes as compared to that of electrons in GaAs, 
LLM is enhanced in p-doped samples as compard to n-doped samples with the same density \cite{santos92}.
We report the first unambiguous observation of a FQHS in the second LL
of a 2DHS at $\nu=8/3$. This was possible because of the combination of
progress in the growth of exceptional quality Carbon-doped 2DHS \cite{manfra05,gerl05}
and of achievement of ultra low charge carrier temperature \cite{csa11}.
The $8/3$ FQHS has an energy gap of 40~mK and, to our surprise, 
its particle-hole symmetric pair at $\nu=7/3$ 
does not develop down to the lowest temperatures of 6.9~mK. 
This observation is contrary to that in electron samples where the $\nu=7/3$ FQHS is typically more robust than the $\nu=8/3$ FQHS. Our data shows that the absence of the $7/3$ state down to the lowest temperatures reached
is unlikely to be caused by a spin transition and we conclude, therefore, that it is most likely a LLM effect. 


The two samples used in this study were cleaved from the same wafer, which is a Carbon-doped 20nm wide GaAs/AlGaAs quantum well grown on the high symmetry surface (100) of GaAs. 
We chose a Carbon-doped 2DHG grown on (100) over Si-doped samples grown onto (311)A because of
a simpler band structure, a more isotropic conduction in the absence of a magnetic field, and because
of superior hole mobilities achieved at similar densities
\cite{manfra05,gerl05,santos92b}.
After illumination with red light the first sample had a density of $6.2\times 10^{10}$~cm$^{-2}$ and mobility $2.7\times 10^6$~cm$^2$/Vs at 6.9~mK. The second sample was thinned down to 100~$\mu$m in order to change the carrier density by backgating. 
Eight Ohmic-contacts were prepared on the perimeter of these 4~mm$\times$4~mm square pieces from In/Zn alloy.

\begin{figure}[b]
 \includegraphics[width=1\columnwidth]{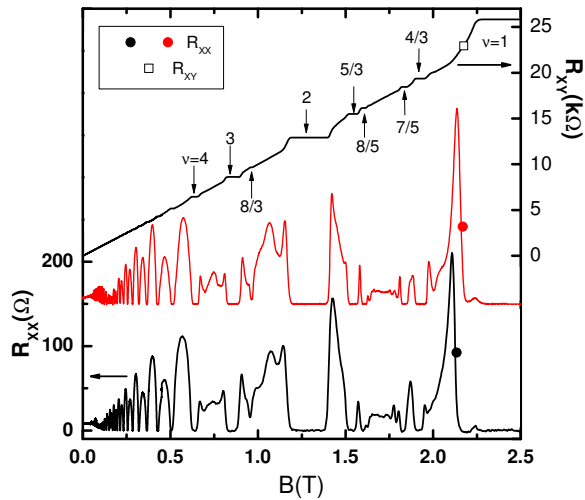}
 \caption{\label{f1}
Magnetotransport data of the ungated sample at 6.9~mK. The two R$_{xx}$ traces are measured along perpendicular directions and show the absence of a strong anisotropy even at finite $B$-fields. 
}
\end{figure}

Magnetotransport measurements at ultra-low temperatures were performed at an excitation of 2~nA in a custom designed Oxford-400~$\mu$W dilution refrigerator. 
At mK temperatures poor thermal contact often results in a saturation of the effective charge carrier temperature at a value higher than that of the fridge.
In order to mitigate this the sample was soldered onto sintered Silver electrodes which were immersed into a liquid He$^3$ bath \cite{xia04,csa11}. 
Temperature is measured by monitoring the magnetic field independent viscosity of the He$^3$ with a quartz tuning fork 
immersed into the same He$^3$ bath \cite{csa11}.
Since we cannot measure the temperature of the charge carriers directly, we monitor a transport feature which depends strongly on $T$. 
For this purpose we have chosen the $\nu=2$ integer quantum Hall state shown in Fig.1. As seen in Fig.2b, the width of the $\nu=2$ plateau does not saturate but changes very rapidly instead with decreasing $T$. We therefore believe that the temperature of our charge carriers follows that of the He$^3$ bath to the lowest temperatures. 

Fig.1 shows the longitudinal resistance R$_{xx}$ and Hall resistance R$_{xy}$
of the ungated sample at a bath temperature of $T=6.9$~mK. The terminal filling factor at the largest $B$-fields is $\nu=1/3$ (not shown), the same as in 2DHS grown on the (311)A surface \cite{santos92}. The energy gap at $\nu=1/3$ 
$\Delta_{1/3}=1.74$~K exceeds by 16\% the value 1.5K reported in
a 2DHS with a similar density $6.55 \times 10^{10}$~cm$^{-2}$ grown on (311)A surface  
\cite{santos92b} demonstrating the exceptionally high quality of this sample. 
We also observe a large number of fully quantized FQHS in the lowest LL such as the ones at $\nu=4/3$, $5/3$ \cite{mano94b,muraki99,gerl05,manfra07}, 
and, for the first time, at $\nu=7/5$, $8/5$.

\begin{figure}[t]
 \includegraphics[width=.9\columnwidth]{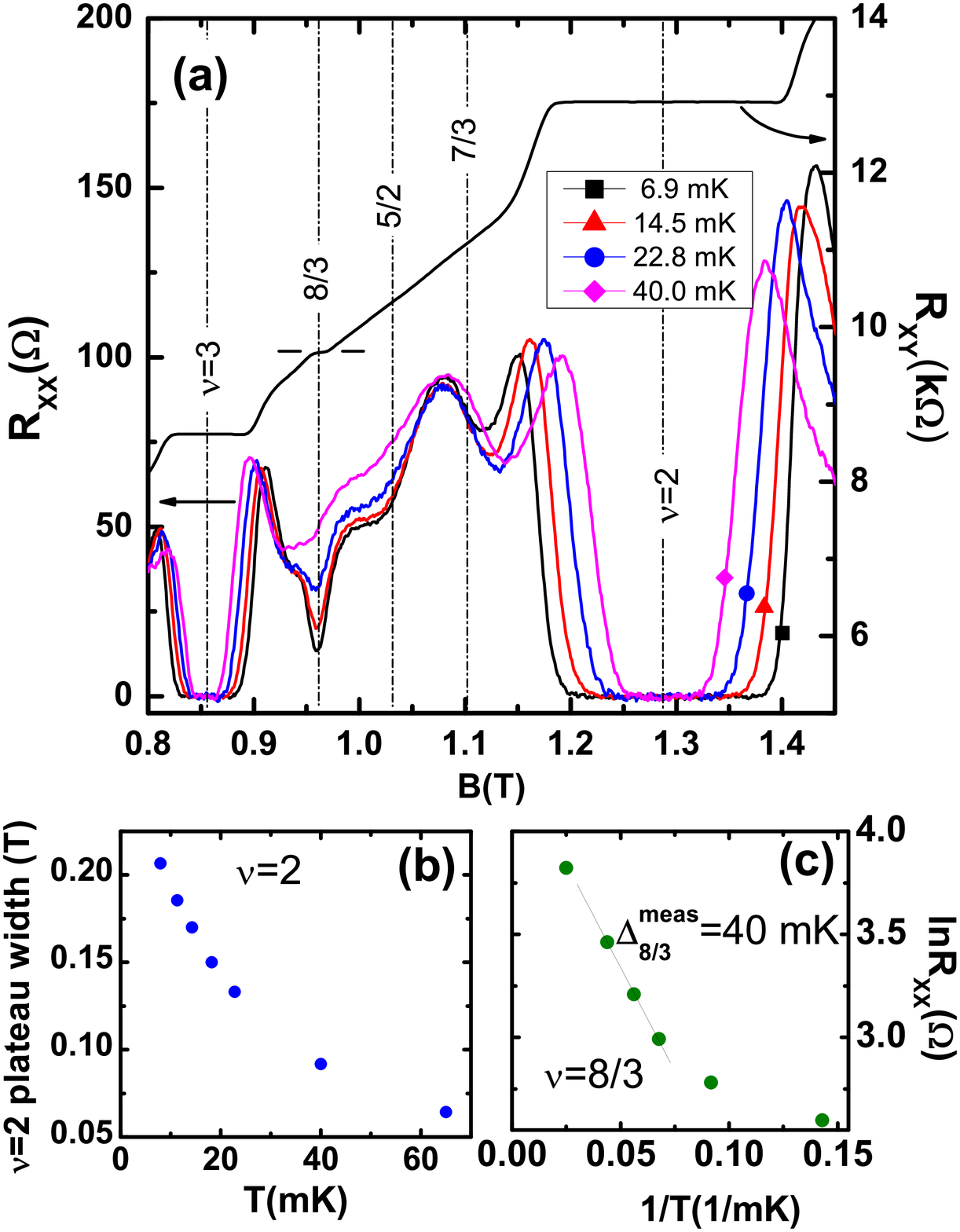}
 \caption{(a) $T$-dependence of R$_{xx}$ and R$_{xy}$ at 6.9~mK in the second LL region. 
 (b) The $T$-dependence of the width of the $\nu=2$ plateau. 
 (c) $T$-dependence of R$_{xx}$ at $\nu$=8/3 shows an activated behavior. 
 \label{Fig2}}
 \end{figure}
 
Fig.2a shows details of the second Landau level transport between filling factors 2 and 3. We observe
a well developed FQHS at $\nu=8/3$ signaled by a Hall plateau quantized better than
0.2\% as referenced to the $\nu=2$ plateau. The Arrhenius plot of Fig.2c reveals an activated behavior with a gap $\Delta_{8/3}=40$~mK.
The deviation from the activated law at the lowest $T$ seen in Fig.2c is 
commonly reported in transport data and it is thought to be due to a change from thermally activated 
conduction to hopping. In our sample we do not observe any features at 
$\nu=5/2$ \cite{mano94,manfra07}
and, unlike in higher density Carbon-doped 2DHS, R$_{xx}$ at $\nu=7/2$ and $11/2$ is isotropic \cite{manfra07}.
 
In Fig.2a we also observe a broad minimum in R$_{xx}$ centered around 1.13~T but this minimum is not accompanied by any discernable features in R$_{xy}$ and therefore we conclude it is not a signature of a FQHS at $\nu=7/3$. Another broad feature in R$_{xx}$ with no signature in R$_{xy}$ is also seen at $\nu=8/3$ above 40~mK, a temperature at which the $\nu=8/3$ FQHS does not survive. A similar broad feature in R$_{xx}$ at $\nu=8/3$ has been reported in Ref.\cite{rodgers93} at 100~mK and in Ref.\cite{davies91} at 50~mK in a tilted $B$-field, but without mentioning a quantized R$_{xy}$ plateau or an activated transport and therefore those features can hardly be ascribed to a FQHS. We thus report the first unambiguous observation of a FQHS in the second LL of a 2DHS at $\nu=8/3$.

It is remarkable that the $\nu=8/3$ FQHS develops at the very low $B$-field of 0.96~T at which
no FQHS of the second LL has been seen in either 2DHS or 2DES. Hence the possibility of a spin transition has to be considered which is identified by a gradual decrease followed by an increase of the gap as either the in-plane $B$-field or the density is varied \cite{rodgers93,davies91,eisen90}. In order to avoid possible anysotropic stripe phases induced by tilted field observed in
2DES \cite{xia10} we investigated the response of the states to backgating.
In spite of the $\nu=8/3$ FQHS being adversely affected by the degrading of the sample due to processing we still discern an inflexion point in R$_{xy}$. As seen in Fig.3, in the $8.77$ to $5.15\times 10^{10}$/cm$^2$ density range we do not observe a strengthening of the $\nu=8/3$ FQHS or an emergence of the $\nu=7/3$ FQHS. Thus in the density range accessed we do not observe a spin transition for either the $8/3$ or the $7/3$ FQHS. 

The effective mass of 2D holes in GaAs can be larger by a factor of 5
as compared to that of electrons. As a consequence, 
LLM is enhanced by the same factor in 2DHS as compard to 2DES at a given 
density \cite{santos92}. The strength of the LLM is encoded into the LLM parameter
$\kappa$ defined as the ratio of the Coulomb and cyclotron energies \cite{yoshi86}. 
Using an effective mass m$_{eff}=0.39$m$_e$ for our Carbon-doped 2DHS\cite{zhu07} we find $\kappa=14.8$ 
at $\nu=8/3$. This value is one order of magnitude larger than
$\kappa=1.6$, the largest LLM parameter at which $\nu=8/3$ FQHS has been previously
reported in 2DES \cite{dean08,nuebler10}. Thus the $\nu=8/3$ FQHS in our 2DHS develops in the limit of extremely strong LLM. 

By ruling out the possiblility of a spin transition in the density range accessed
we surmise that the different relative strength of the $8/3$ and $7/3$ FQHS in electron and hole samples 
is likely caused by LLM. LLM is known to break particle-hole symmetry
\cite{levin07,lee07,wan08,wang09,peterson08,wojs10,bishara09,rezayi11,wojs10}
and it might change the relative strength of the 7/3 and 8/3 FQHS. 
A known example of particle-hole asymmetry for FQHS in the second LL thought to be induced by LLM
is that of the well developed $\nu=12/5$ but missing particle-hole conjugate $\nu=13/5$ FQHS in electron samples 
\cite{xia04,choi08,csa10}.

The well developed FQHS at $\nu=8/3$ together with no transport signature at $\nu=7/3$ in our 2DHS 
is in stark contrast to the observations in 2DES in which the gap 
of the $7/3$ FQHS is found to be larger than that of the $8/3$ FQHS 
in the vast majority of reports
\cite{pan99,eisen02,miller07,dean08,pan08,nuebler10,zhang10,csa10}. We know of only one exeption \cite{choi08}. 
We have argued that this unexpected violation of the particle-hole symmetry in the second Landau level in our 2DHS
must be due to LLM.
In the absence of a thorough understanding of the details of particle-hole symmetry breaking the nature of the $8/3$ FQHS in our 2DHS remains unresolved but our data hints toward a possible change of the nature of the
8/3 and/or 7/3 FQHS with increasing LLM.

\begin{figure}[b]
 \includegraphics[width=.9\columnwidth]{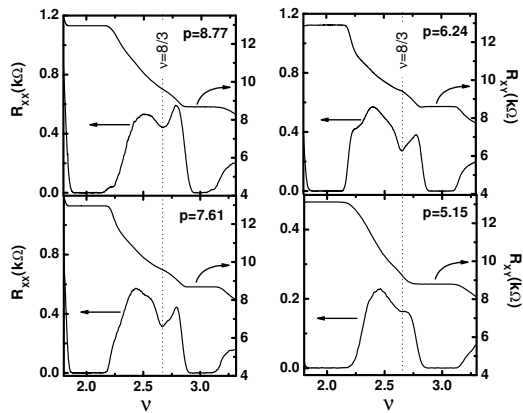}
 \caption{Density dependence of the magnetoresistance of the backgated sample at 6.9~mK. Densities are indicated in units of $10^{10}$cm$^{-2}$.
 \label{Fig3}}
 \end{figure}

In summary, we found for the first time a quantized FQHS at $\nu=8/3$ in the second LL and at $\nu=7/5$ and $8/5$ in the lowest LL of a 2DHS.  An interesting unexplained feature of our data is the absence of the $7/3$ FQHS which we conjecture is a result of the particle-hole symmetry breaking effects due to strong LL mixing. Our results in the 2DHS together with results in 2DES point towards a need of theoretical models which include such symmetry breaking terms.

We thank L. Rokhinson for sharing his software and for Y. Lyanda-Geller for stimulating discussions. N.S. and G.A.C. were supported on 
NSF grant DMR-0907172, M.J.M. acknowledges the Miller Family Foundation, and L.N.P. and K.W.W. 
the Princeton NSF-MRSEC and the Moore Foundation.

\bibliography{FQHE}

\end{document}